\documentclass[showpacs,preprintnumbers,prl,preprint]{revtex4}

\usepackage{amsfonts}
\usepackage{amsmath}
\usepackage{graphicx}
\usepackage{dcolumn}
\usepackage{bm}

\begin{document}

\preprint{}
\title{Quantitative detection of electrically injected spin accumulation
in GaAs using the magneto-optical Kerr effect}
\author{A.~V.~Kimel$^{1}$, A.~Kirilyuk$^{1}$,
P.~Grabs$^2$, F.~Lehmann$^2$, A.~A.~Tsvetkov$^1$, G.~Schmidt$^{2}$,
L.~W.~Molenkamp$^{2}$, and Th.~Rasing$^{1}$} 
\affiliation{$^{1}$Institute for Molecules and Materials,
Radboud University Nijmegen, 6525 ED Nijmegen, The Netherlands\\
$^{2}$Physikalisches Institut der Universit\"{a}t W\"{u}rzburg, 97074
W\"{u}rzburg, Germany}
\date{\today }

\begin{abstract}
Using an ultra sensitive magneto-optical Kerr rotation setup we have observed
electrical spin injection from (Zn,Mn)Se into bulk GaAs and quantified the spin
injection efficiency. The current induced contribution was carefully separated
from possible artifacts and studied as a function of voltage and external
magnetic field. Our measurements allow us to estimate the concentration of the
electrically injected spins into GaAs to be approximately
0.4$\times10^{15}$cm$^{-3}$ at a reverse bias of 0.7~$V$.
\end{abstract}

 \maketitle

The study of spin-dependent properties of semiconductors has become one of the
hottest topics in physics due to their importance for the newly developing area
of ``semiconductor spintronics'' \cite{kim1,kim2}. The main idea of
``spintronics'' is to use the fact that electrons have not only a charge but
also a spin, yielding an additional degree of freedom for electronic devices.
The realization of semiconductor spintronic devices, however, requires
efficient electrical spin injection into a semiconductor, a task which is far
from trivial and subject to intense research activity
\cite{kim3,kim4,kim5,kim6,kim7,kim8,kim9,kim10}. Besides, to control and
switch the magnetization using spin-polarized currents \cite{torque,torque2}, 
the spin accumulation at electrode/magnetic-layer interfaces is a crucial
parameter \cite{slon}, that however is hard to access experimentally. 

Most optical studies of electrical spin injection into semiconductors so far
were based on photoluminescence measurements. However, this method has
considerable limitations regarding sample structure, quality, and spectral
range \cite{kim3,kim4,kim5,kim6,kim7,kim8,kim9,kim10}. Moreover, the study of
electrically injected spins based on photoluminescence can only be used to
determine the presence or absence of spin-polarization of the injected current,
while the spin accumulation or the number of injected spins is not accessible.
The magneto-optical Kerr effect (MOKE), however, can be used to overcome all these
limitations. Using MOKE spin injection and accumulation 
can be studied with no limitation with
respect to sample and in a broad spectral and thermal range, including
wavelengths were no luminescence can be measured \cite{kim11,kim12}. In this
letter, we use a (Zn,Mn)Se/GaAs heterostructure to demonstrate that the
magneto-optical Kerr effect is a powerful tool for the quantitative study of
electrical spin injection. At a reverse bias of -0.7 V we observed a maximum
spin-injection from a (Zn,Mn)Se electrode into GaAs of about
0.4$\times10^{15}$cm$^{-3}$.

The magneto-optical Kerr effect is rotation of the polarization of light upon
reflection from a magnetized sample. In terms of dielectric permittivity tensor
$\epsilon$ the Kerr rotation $\theta_k$ can be presented as
\begin{equation}
i \theta_k = \frac{2 \epsilon_{xy}}{(n+1)^2 n} \ ,
\end{equation}
where $n$ is the refractive index of the sample. 
It can be shown from the Onsager's principles that only off-diagonal components
of $\epsilon$ can be odd functions with respect to the
magnetization of a medium. Neglecting terms of the higher order one can assume
$\epsilon^{(a)}_{ij} \approx \alpha \cdot M$, and re-write equation (1) in
the following form:
\begin{equation}
i \theta_k = \frac{2 \alpha M}{(n+1)^2 n}
\end{equation}
Normally for dia- and paramagnetic materials the magnetization of the medium is
created by an external magnetic field. However, in the presence of spin
injection, magnetization can be created by an nonequilibrium spin polarization
as well. Thus, the magneto-optical Kerr effect can be a sensitive probe of the
artificially introduced spin polarization in solids in general, and in
semiconductors in particular.

The structure under investigation consisted of a paramagnetic 
(Zn,Mn)Se spin aligner on a conducting GaAs layer in an applied magnetic field. 
This field will Zeeman-split the conduction band levels in (Zn,Mn)Se leading
to a, in principle 100\%, spin polarized injection current from the
(Zn,Mn)Se electrode into GaAs. 
The structure was based on a highly n-doped GaAs
substrate onto which a n-type GaAs layer with a doping of n=4$\times
10^{16}$cm$^{-3}$ and a thickness of 400 nm was deposited by molecular beam
epitaxy. After deposition, the sample was transferred into a second growth
chamber without breaking the UHV conditions. In this chamber a 100 nm
(Zn$_{0.89}$Be$_{0.06}$Mn$_{0.05}$)Se layer was deposited which was also n-doped
(approximately $n=4\times 10^{18}$cm$^{-3}$). In order to achieve a good ohmic
top contact another 30 nm thick ZnSe layer with a n-type doping of $n=2\times
10^{19}$cm$^{-3}$ was deposited as a top layer. Again without breaking the UHV,
the sample was transferred to a metallization chamber in which a metallization
of Al, Ti, and Au was deposited by e-beam evaporation with layer thicknesses of
10 nm, 10 nm, and 30 nm, respectively. On top of this metallization a
rectangular frame-like contact of 10 nm Ti and 250 nm Au was fabricated by
optical lithography and lift off. Subsequently, Au and Ti inside the frame
were removed by dry etching, leaving a transparent, but conducting window
for the MOKE measurement. The layer sequence in the structure is virtually the
same which was used for spin injection into a light emitting diode in
\cite{kim5}. It was used again in the present experiment because it typically
yields a spin polarization of the injected current of up to 90 \%.

For the MOKE measurement we used an ultrasensitive laser polarimeter. The light was
incident from the side of the (Zn,Mn)Se electrode. After transmission though
this spin injector the light was reflected from the (Zn,Mn)Se/GaAs interface.
For the excitation a wavelength of 810 nm was chosen, because (Zn,Mn)Se is
transparent at this wavelength, whereas the absorption in GaAs is quite large,
yielding a maximum reflection at the (Zn,Mn)Se/GaAs interface (see inset in
Fig. 1). Moreover, the photon energy at this wavelength is close to the exciton
resonance in GaAs and all magneto-optical effects in this semiconductor of interest 
are resonantly enhanced. The sample was placed in a static magnetic field directed
along the growth direction of the sample aligning the spins in the (Zn,Mn)Se in
the direction of the magnetic field. Rectangular current pulses with a
repetition frequency of 400 Hz were sent through the structure and the rotation
of the polarization of the reflected light was detected at the same frequency.

It should be noted that electrical injection of spins assumes the application
of a voltage that would create relatively high electric fields in the studied
structure, which, in turn, can also modify the dielectric permittivity tensor.
Thus if one uses the magneto-optical Kerr effect for the study of the
electrically injected spin polarization, a careful analysis of the observed
effects and separating them from artifacts is necessary. Regarding the
possible artifacts, the off-diagonal component of the dielectric permittivity
tensor can be written in the following form, neglecting the higher-order terms:
\begin{equation}
\epsilon^{(a)}_{ij} = \alpha M + \beta EH + \gamma I + ...\ ,
\end{equation}
where $H$ is the external magnetic field, $E$ is the electric field and $I$ is
the current. The off-diagonal component of the dielectric permittivity tensor
can not be proportional to $IH$-component since it violates the Onsager's
principle. One can see that in case of spin injection into GaAs, the
induced magnetization should be a linear function of the electrical current
through the structure, as well as of the external magnetic field, since the
latter affects the alignment of the spins in ZnMnSe, and thus determines spin
polarization of the current. In principle, the term $EH$ in Eq.\ (3) can
manifest itself in the magneto-optical Kerr effect in a way similar to the
electrically injected spin polarization. However, these two contributions into
the Kerr effect can be clearly separated in the structures with a nonlinear
$I/V$-characteristics.

\begin{figure}[tbp]
\includegraphics[width=.46\textwidth]{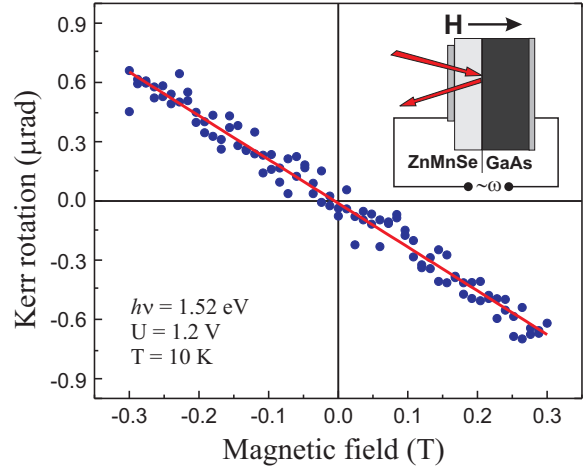}
\caption{MOKE signal at an applied voltage of 1.2 eV and a photon energy of
1.52 eV as a function of the external magnetic field. The inset shows the
geometry of the experiment.} \label{fig:1}
\end{figure}

Fig.\ 1 shows that the MOKE signal, induced by a spin polarized current in the
(Zn,Mn)Se/GaAs heterostructure, depends linearly on the external magnetic
field. This linearity originates from the fact that the spin injector (Zn,Mn)Se
is paramagnetic. For magnetic fields used in our experiment, the paramagnetic
susceptibility is far from saturation and the magnetization as well as the
spin-injection efficiency are approximately proportional to the external
magnetic field.

Fig.\ 2 shows the MOKE signal as a function of the current applied to the
structure. One can see that for positive bias, i.e. current direction from the
GaAs into (Zn,Mn)Se, no MOKE signal is detected. However, if the bias is
negative and current flows from (Zn,Mn)Se into GaAs, a pronounced MOKE
signal is observed with a well distinguished maximum at 0.7 V.

\begin{figure}[tbp]
\includegraphics[width=.46\textwidth]{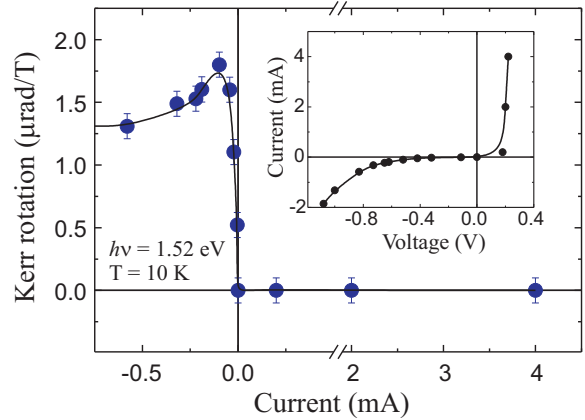}
\caption{MOKE signal as a function of the amplitude of the current pulses at a
photon energy of 1.52 eV with an external magnetic field equal to 0.3 T. The
inset shows the $I/V$-characteristics of the structure.} \label{fig:2}
\end{figure}

The observed current dependence of the MOKE signal is in excellent agreement
with the theoretical dependence of spin accumulation on injection current
\cite{kim15}. Only a minor spin accumulation is expected when the carriers flow
from the GaAs into the (Zn,Mn)Se. In contrast, when the electrons flow from
(Zn,Mn)Se into GaAs, spin polarized carriers of (Zn,Mn)Se are injected into the
GaAs, causing a non-equilibrium spin accumulation in the latter. The current
dependence of this spin accumulation can be readily explained by taking into
account several effects. At low bias transport occurs in the regime of linear
response and consequently the spin accumulation is approximately proportional
to the current. In this regime the spin accumulation can be described simply by
the diffusion equation as in \cite{Schmidtrap}. With increasing current the
spin accumulation leads to a band bending at the (Zn,Mn)Se/GaAs interface which
reduces the spin polarization because of a redistribution of spins between the
two spin bands in the (Zn,Mn)Se \cite{kim14}. At the current level where band
bending starts to occur (at 0.7 $V$), a maximum of the induced spin accumulation is reached
and thus a maximum in the MOKE signal is observed. For still higher bias, one
is in the drift regime \cite{Yu02} where spin transport is characterized by an
extension of the effective spin diffusion length due to the unscreened electric
fields in the device. This leads to an increased penetration depth of the spin
accumulation in the GaAs. However, once this non-linear regime is reached, the
magnitude of the spin accumulation remains constant  and hence the MOKE signal
also does not increase further. These observations clearly demonstrate that the
detected MOKE signal is indeed induced by the spin polarized current injected
from (Zn,Mn)Se into GaAs, and not by the term proportional to $EH$. Thus the
magneto-optical Kerr effect is a measure of the efficiency of spin injection
into semiconductors.

In order to estimate the efficiency of the electrical spin injection we used
the optical orientation of the electron spins in GaAs as a reference. It is
well known that under influence of circularly polarized light, a
non-equilibrium spin polarization can be created in a semiconductor. For
semiconductors with zinc-blend structure, such as GaAs, this phenomenon is well
studied \cite{orient}. Therefore knowing the intensity and wavelength of the
photoexcitation, one can estimate the number of photoinjected spin polarized
carriers. Note that such an estimate also requires a detailed knowledge of the
probabilities of the photoexcited transitions and the absorption coefficient of
GaAs, but these are textbook values \cite{ioffe}.

For our calibration experiment we have created spin polarized carriers by
pumping with circularly polarized light and probed the resulting spin
polarization with the magneto-optical Kerr effect in a pump and probe
configuration. Details of the experiment were described elsewhere \cite{kim11}.
In brief, for the measurements a pulsed Ti:sapphire laser was used, with a
pulse duration of approximately 100 fs and a repetition rate of 82 MHz, with a
photon energy of 1.52 eV. Coherent pump and probe pulses with an intensity
ratio of 10:1 were focused on the sample. The pump fluence on the sample was of
the order of 10 $\mu$J/cm$^2$ and produced a spatially averaged nonequilibrium
spin concentration of about $0.7\times10^{17}$cm$^{-3}$. The Kerr rotation
experienced by the reflected delayed probe pulse is measured as a function of
the time delay between pump and probe pulses.

Fig.\ 3 shows the temporal behavior of the magneto-optical Kerr effect as a
function of time delay between pump and probe pulses. It is well known that the
photo-induced magneto-optical Kerr effect is sensitive not only to spin but
also to carrier dynamics. Nevertheless, at a time delay of 15 ps one may expect
that processes related to carrier dynamics are completed, while the spin
relaxation of the photoexcited electrons is expected to be much longer. Thus
the value of the Kerr rotation at a time delay of 15 ps can be taken as the one
that corresponds to a spin concentration $n_s = 0.7\times10^{17}$cm$^{-3}$. One
can see from Fig.\ 3 that in GaAs at a photon energy of 1.52 eV and at a
temperature of 10 K the spin concentration of $n_s = 0.7\times10^{17}$cm$^{-3}$
results in a Kerr rotation of about 10$^{-4}$ rad. This ratio can be used for
the calibration of the MOKE detection of electrically injected spins into bulk
GaAs. Using this ratio between the Kerr rotation and spin concentration one can
estimate that at conditions that corresponds to the maximum efficiency of the
spin injection in our experiment ($U=-0.7$~V and $H = 0.3$~T) the concentration
of the electrically injected spins is about $0.4\times10^{15}$cm$^{-3}$.

\begin{figure}[tbp]
\includegraphics[width=.46\textwidth]{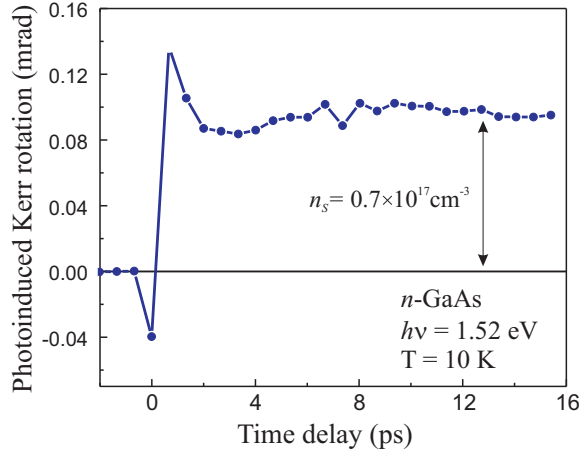}
\caption{The photoinduced magneto-optical Kerr effect as a function of the time
delay between pump and probe pulses. At a time delay of 15 ps the Kerr effect
is assumed to be due to the nonequilibrium spin polarization.} \label{fig:3}
\end{figure}

In summary, using the example of the heterostructure (Zn,Mn)Se/GaAs we have
demonstrated that the magneto-optical Kerr effect is a powerful tool for the
study of electrical spin injection. We have shown a current induced
contribution to the magneto-optical Kerr effect that arises from spin polarized
carriers electrically injected from (Zn,Mn)Se into GaAs. This current induced
contribution was carefully separated from possible artifacts and allowed us to
estimate that the concentration of the electrically injected spins into GaAs is
about $0.4\times10^{15}$cm$^{-3}$ for a voltage of 0.7 V and a magnetic field
of 0.3 T.

\medskip

This work  was supported in part by the EU RTD project SPINOSA, the EU Network
DYNAMICS, the Nederlandse Organisatie voor Wetenschappelijk Onderzoek (NWO) as
well as Stichting voor Fundamenteel Onderzoek der Materie (FOM), Deutsche
Forschungsgemeinschaft (DFG) and the DARPA Spins program.


\begin{references}
\bibitem{kim14} Note that the linearity cannot be inferred from
the occupation of the Zeeman levels but stems from non-linear spin injection
effects which occur at the elevated voltages which are used in the experiment
and which were described in \cite{kim15}
\end{references}
\end{document}